# Extracting preseismic electric signals from noisy Earth's electric field data recordings. The "noise injection" method.

Thanassoulas[1], C., Klentos[2], V., Verveniotis, G.[3]

1. Retired from the Institute for Geology and Mineral Exploration (IGME), Geophysical Department, Athens, Greece.
   e-mail: thandin@otenet.gr - website: www.earthquakeprediction.gr

2. Athens Water Supply & Sewerage Company (EYDAP),
   e-mail: klenvas@mycosmos.gr - website: www.earthquakeprediction.gr

3. Sub-Director, Physics Teacher at 2nd Senior High School of Pyrgos, Greece.
   e-mail: verveniotis_ge@hotmail.com - www.earthquakeprediction.gr

**Abstract.**

An entirely different approach is used for the extraction of preseismic electric signals from highly contaminated by noise data series. The basic element of this method is the "Dirac Delta Function". Instead of applying any known method of filtering, noise is injected, in the data series at different amplitude levels (**p** value) and the generated "family" of filtered data series is compared to different convergence criteria. In the case of preseismic electric signals identification the most appropriate (**p**) value is selected by testing the convergence of generated intersections of more than three "families" of filtered data which were generated from more than three monitoring sites. The "noise injection" method was tested against real data recorded long before two large earthquakes in Greece and one in Turkey. The obtained results justify the validity of the postulated method.

## 1. Introduction.

Before any use is made, of the data series which result from the acquisition and registration of the Earth's electric field, by any monitoring site, two important operations are applied on them. The first one is the editing of the data and the second is the rejection of the superimposed noise.

The noise rejection (filtering) techniques and methodologies, which deal with time series of any data type, generally require, as input data, files which are free from data gaps. If, accidentally, a "gap" is met during the processing of a data file, then in most cases, the running procedure, either "crashes" or generates erroneous results. In both cases, it takes some time to be wasted, in order to rectify this nasty situation. Therefore, as a preliminary step, before any methodology is applied on a specific data set, it is a good policy to check against data gaps and to apply any suitable methodology to recover the data continuity.

In the particular case of the registration of the Earth's electric field at the various monitoring sites which are in operation (www.earthquakeprediction.gr) to date **(ATH, PYR, HIO),** the following main causes have created data gaps:

   a. Damage on the receiving dipole electrode lines. This is mostly a breakdown (due to various causes) of the wires which connect the electrodes with the pre-processing unit, located at the housing of the monitoring site.

   b. "Crashing" of the used computer system, due to power line voltage rapid and large amplitude changes, which the used ordinary UPS, cannot accommodate.

In both cases, the result is the same. Data gaps are created, since the operator in charge of the monitoring site becomes aware of the faulty situation, only after a few hours from the data gap occurrence.

As long as such a data gap has been detected, the gap is replaced by linearly, interpolated data, taking into account the start and the end data values, which preceded and followed the data gap. This procedure is presented in the following figures **(1 – 3)** which present data gaps, met at recordings generated by **PYR** monitoring site.

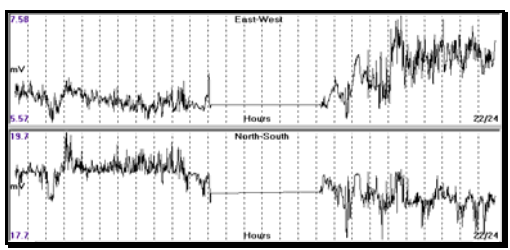

Fig. 1. Presentation of "missing data" and their linearly, interpolated values, used to bridge the corresponding data gap. Date of gap recording is 14th November 2006, at Pyrgos **(PYR)** monitoring site.

The observed data gap of figure **(1)** lasted for six (**6**) hours during the actual recording of 14th November, 2006 at Pyrgos **(PYR)** monitoring site. A much longer data gap is presented in the following figure **(2)**.

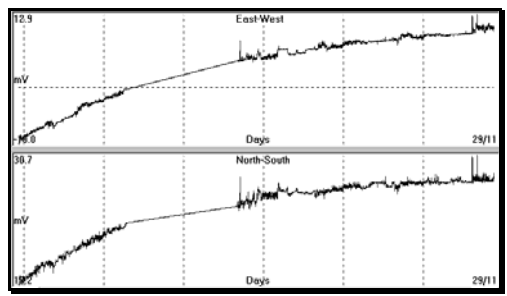

Fig. 2. "Missing data" and their linearly interpolated values are presented, which are used to bridge the corresponding data gap. Date of gap recording is 25th - 26th November 2006, at Pyrgos **(PYR)** monitoring site.

The observed data gap of figure **(2)** lasted for more than a day (25[th] – 26[th] November 2006). Finally a third example is presented in figure **(3)**.

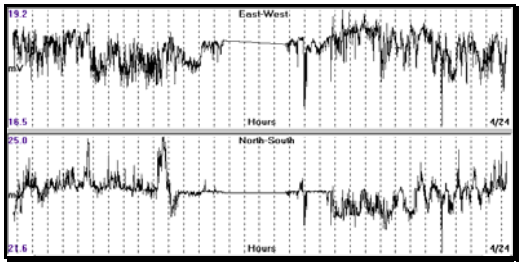

Fig. 3. "Missing data" and their linearly, interpolated values are presented, which were used to bridge the corresponding data gap. Date of gap recording, 16[th] January 2007, at Pyrgos **(PYR)** monitoring site.

The procedure, which is followed, in order to fill the data gap, does affect the "short wave length", electrical signals, recorded, with period of some hours at most. In practice, this data gap time period is replaced by a "noise free", new data set. However, it does not change the character of the entire data set, in the long wave length domain and therefore, does not affect the procedures which are applied for the epicenter area determination which deal, mainly, with much more longer periods (Thanassoulas, 2007).

Consequently, the error, which is introduced, on the data set by this type of "data preprocessing", can be considered as negligible.

## 2. Preseismic electrical signal identification.

The way for obtaining a continuous in time data set from the registration of the Earth's electric field, which is observed on ground surface, has been so far demonstrated. More over, this data set can be subjected to "pre-whitening operations" or any other type of filtering, so that it results in to a "clean" recording. So far, no assumption was made for the type of anomalous signals content in any data set. Therefore, any kind of signals such as of anthropogenic origin, of ionospheric origin and of seismogenic – tectonic origin could be incorporated in any "cleaned and continuous" data set.

Consequently, the question, which arises, is: does a particular data set contain any precursory, electrical signals and if so, how is it possible to identify them?

It is not an easy task to answer this question. Actually, it takes knowledge of the insight of the physical mechanisms, which are activated each time during the preparation stage of an earthquake, which unfortunately, is not clearly known to the researchers, yet. Even if these physical mechanisms were known, still the problem exists because of the variability of the mechanisms, triggered each time, in the same seismogenic area, and at different time periods. Consequently, the signature of the preseismic, electrical signal, to be generated, is still unknown.

Many researchers studied this specific problem, of the identification of the preseismic, electrical signals. Varotsos and Lazaridou (1991) used the **dV/dl = constant** criterion to compare potential values, registered, with short dipoles of different length at the same monitoring site; the preseismic, electric signal must appear simultaneously on the short and long dipoles of the stations, concerned; the magnetotelluric (**MT**) noise can be distinguished in the recordings, since it is recorded simultaneously in all dipoles, in all monitoring sites; while, "the polarity and amplitude of the preseismic signal on the short and long dipoles must be compatible with a distant source assumption". More over, guidelines and examples were presented for the discrimination of preseismic, electrical signals from artificial noise and from magnetotelluric changes (summary in: Varotsos, 2005). However, the later criteria were strongly objected by Tzanis and Gruszow (1998).

The same problem was tackled, in a different approach, by Ozima et al. (1989). The Bayesian approach (Ishiguro 1981; Ishiguro et al. 1984) of the analysis of the data, related to crustal movements, was adopted and the observed Earth potential data were separated into four components: tidal, electromagnetically induced, trend and irregular component.

Thanassoulas (1991) used traditional Fast Fourier Transform (**FFT**) in order to identify very long period (**VLP**) signals, incorporated in the recordings of the Earth's electric field, signals which were registered by the **VAN** group prior to Kalamata earthquake (15[th] September 1986, Ms = 6.2R). In this work, anomalous signals of period less than a day were considered as "noise" of any kind and therefore, were completely rejected by the use of a low pass filter. The output of this filter was used, successfully, to determine the azimuthal direction of Kalamata EQ, in relation to the location of the recording station.

Cuomo et al. (1996) proposed a statistical method, to evaluate extreme events in earthquake precursory signals of electric nature. The occurrence probability of seismo-electrical anomalies is computed by means of a parametric time series approach. In a second step, an autoregressive model, to describe the residual time series, is suitably identified and fitted to the data. Finally, parametric inference of extreme events is carried out on the basis of the selected model.

Enomoto et al. (1997) used a different approach for the characterization of pulse-like, geoelectric signals as being related to seismic activity. In this methodology the anomalous electrical signals are compared in turn with: a certain threshold level, simultaneous recording by more than one stations, presence of lightning, association of thunder clouds or radar echo, simultaneous presence of the signal in the VLF band. In a positive result, the signal is considered as a possible seismic precursor and is compared to any seismic activity at a threshold level of 5.5R.

Cuomo et al. (1998) investigated the time dynamics of geoelectric, precursory time series, using autoregressive models. The approach he uses allows information, to be obtained, for the geophysical system, which produces the electrical phenomena in seismic, active areas, when the only information about the time series comes from the time series themselves. It is based on two forecasting approaches: the global, autoregressive approximation and the local, autoregressive approximation. The first approach views the data as a realization of a linear, stochastic process, whereas, the second approach considers the data points as the result of a deterministic process, supposedly nonlinear. The comparison of the predictive skills of the two techniques is a strong test to distinguish between low-dimensional chaos and random dynamics. The later methodology was extended more by the use of Higuchi (1988, 1990) analysis, in order to extract maximum, quantitative information, about the time dynamics from these geoelectric signals (Cuomo et al. 1999). Cuomo et al. (2000) used a slightly different statistical methodology in order to discriminate extreme events in geoelectric, precursory signals with implications on earthquake prediction.

Telesca et al. (2004) applied the Principal Component Analysis (**PCA**) method on geoelectric signals, measured, in the seismically active area of Basilicata region, southern Italy. The analysis showed earthquake precursory patterns in the daily variation of the principal components.

Fisher information analysis was used by Telesca et al. (2005), in order to study the time fluctuations in the dynamics of the geoelectric data, recorded in Tito site, which is located in a seismic area of southern Italy.

Ida and Hayakawa (2006) used fractal analysis for the ULF data, recorded, during Guam earthquake in 1993, to study pre-fracture criticality.

Hayakawa and Timashev (2006) applied flicker noise spectroscopy in an attempt to identify precursors in the ULF geomagnetic data.

Varotsos (2005), after using statistical physics, suggests "all SES signals and activities exhibit critical behaviour, while artificial noise does not". On the basis of criteria of this kind, seismic electric signals (**SES**) are able to be discriminated from artificial noise. However, "artificial noise may some times be associated with criticality (when a system approaches a failure)" and therefore, cannot be discriminated from the **SES**.

In conclusion, to date a large number of methodologies, aiming to discriminate the electric, precursory signals, has been presented in the seismological literature. Each one of them was applied on specific data set and seismogenic area. Therefore, by taking into account the physical complexity of the seismic generating mechanism, it is questionable if these methodologies can be applied in a different seismogenic area and for a different precursory, electric data set, with the same successful results. The obvious question to arise is: what is the solution to this problem? In practice, there must be a universal methodology, which will be applicable to all electrical, precursory data sets. Although the answer will appear to be bizarre, it could be as follows: **all anomalous, electrical signals, no matter their origin is, are initially (as a working hypothesis), considered of being as seismic, precursory signals!!**

In such a case, these signals have to follow simple physical laws of electric fields; such as the electrical field intensity vector should be pointing towards the current source location.

The time data series of electric potential which are registered and edited, so far, consist of any seismic, precursory, electrical signals, if these do exist, contaminated, by industrial or anthropogenic noise, any kind of superimposed electrical trends and of any type of signals, induced, by the ionospheric activity. Therefore, it is very important to separate each one of them from each other, before any use is made for any earthquake prediction attempt.

The procedure which is used for this type of processing is called "filtering". Filtering is distinguished, generally, to "low pass", in which only low frequencies, below a certain value, pass through the used filter, "band pass", in which a certain "frequency band" is allowed to pass through the filter and "high pass", in which only high frequencies, above a certain value, are allowed to pass.

Filtering implementation is made through various well-known methodologies such as:

**- Running a moving average (equally weighted or not).**

This is the most basic scheme, which is used and corresponds to the implementation of a "low pass" filter. In the case of its application in a **differential operator mode** it corresponds to a "high pass" filter. A combined operation of "low-pass" and "high-pass" filters, sequentially, on the same data set, after appropriate selection of the filters characteristics, produces a band-pass output.

**- Polynomial fitting.**

A different approach, for filtering implementation, is made through the use of polynomial fitting to the data set. In this methodology, any data set of **N** equally, spaced data values can be transformed into an **N-1** order, polynomial function. The constant parameter of each polynomial term is evaluated, through a least square (**LSQ**) methodology, from the data values. If the (**Z**) transformation is taken into account, then, by omitting higher order terms of the calculated polynomial, the reconstructed data series corresponds to a new data set that contains only the lower terms of the frequency content of the original data. It is in practice a "low-pass" filtering operation.

**- Fast Fourier Transform (FFT).**

The method, which is mostly used by the scientific community, for any data filtering, is the traditional **Fast Fourier Transform (FFT).** In a very short description of this methodology, the original data are, initially, converted into its frequency spectrum. A filter (of any kind), is used to retain the frequency band of interest, rejecting the entire, unwanted spectrum. Finally, from the retained frequency spectrum, the filtered data, in time domain, are backwards reconstructed. A detailed analysis of the topic was presented by Bath (1974) and Kulhanek (1976).

In the present case of data, which are recorded by the monitoring sites, for the earthquake prediction application, extensive use of the **FFT** methodology is made in order to separate the various oscillating components of the Earth's electric field, which are triggered by the lithospheric, tidal oscillation (Thanassoulas et al., 2001). Below are some examples of the use of the **FFT** procedure on real data.

The first example **(fig. 4)** represents the application of a "low-pass" filter on the time data set of the Earth's electric, **EW** component, registered, by **PYR** monitoring site.

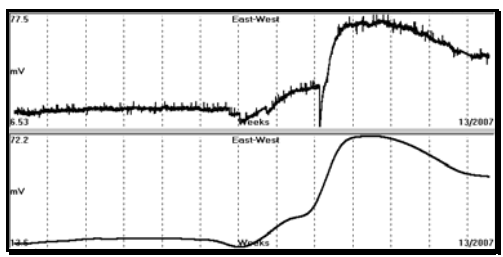

Fig. 4. "Low-pass" filtering of data, recorded by **PYR** monitoring site, for the period 1st January to 28th March, 2007 (**PYR** 070101 – 070328). Stop band of the filter = 10 days (bp=10days). The upper graph indicates the original data, while the lower graph indicates the "low-pass", filtered data.

Furthermore, the "low-pass" filtered data may be assumed as a long wavelength noise or trend. By subtracting this "noise" from the original data, the high frequency content of the registration, is enhanced. Moreover, by applying a "band-pass filter" on the original data, its oscillatory component can be derived. These operations are presented in the following figure **(5)**.

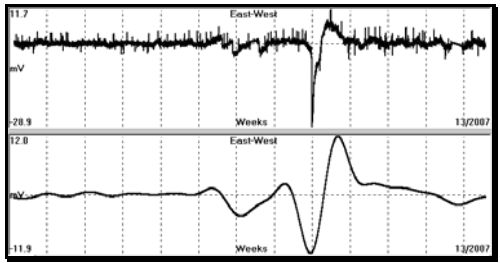

Fig. 5. The upper graph represents the "high-pass", filtered, original data, while the lower one shows the "band-pass" operation with a centre band (cpd = .1) at 10 days and bandwidth (bp = .1) of 10 days, too. The recording period is 1$^{st}$ January to 28$^{th}$ March 2007 (**PYR**, 070101 – 070328).

The same methodology has been applied over a shorter time span of the original recording, but with a different bandwidth. In this case, the chosen parameters are: pass-band center = 1 day, with a bandwidth of half (0.5) of a day. The attempt is to identify, any one-day period, oscillatory component of the Earth's electric field. This is illustrated in the following figure **(6)**.

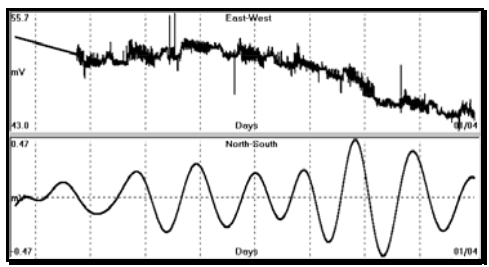

Fig. 6. The upper graph represents the original data, while the lower one the corresponding, oscillatory component of the Earth's electric field, identified, by the use of a "band-pass" filter, with a centre frequency that corresponds to one day's period and a bandwidth of half (0.5) of a day (data recorded by **PYR** monitoring site, **PYR** 070325 – 070401).

The methodology was used for the "pre-whitening" of the original data, in a very similar way. This operation was performed after having chosen appropriate parameters for the filters, so that "white noise" will be eliminated from the original data. The center of the band-pass filter was chosen as 24 cycles/day, while the band-pass of it was set to 12 cycles/day. This operation is demonstrated in the following figure **(7)**.

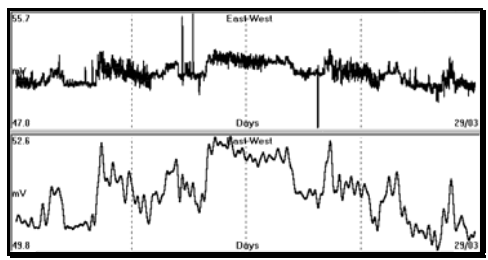

Fig. 7. "Pre-whitening" operation of the original data, recorded, for the period 26$^{th}$ March to 29$^{th}$ March, 2007 by **PYR** monitoring site (**PYR** 070326 – 070329). The centre of "band-pass" filter was set at 24 cycles per day; the bandwidth was set at 12cycles per day.

If the previous results are combined in a differential mode, then the "white-noise" frequency content of the original recording is obtained. This is demonstrated in the following figure **(8)**.

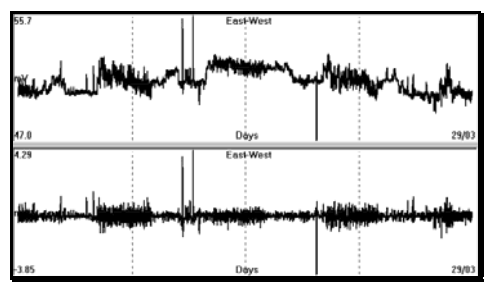

Fig. 8. The upper graph indicates the original data, while the "white noise" content of the original recording is indicated in the lower graph (data recorded by **PYR** monitoring site, **PYR** 070326 – 070329). The centre of band-pass filter was set at 24 cycles per day; the bandwidth was set at 12cycles per day.

The "pre-whitening" procedure was applied on a part of a recording which contains an **SES** signal. The **SES** signal, which was observed, on 4$^{th}$ March, 2007 on the recording of **HIO** monitoring site, is superimposed on an "anomalous" background noise. The "white noise" (and the **SES**, too) was eliminated, so that only the background noise is present. This is demonstrated in the following figure **(9)**.

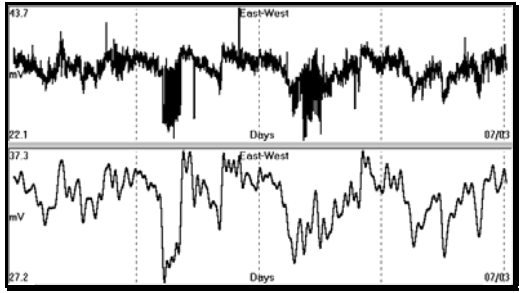

Fig. 9. The upper graph represents the original data with the **SES** present on 4[th] March, 2007, while the lower graph indicates the existing noise (data recorded at monitoring site, **HIO** 070303 – 070306). The centre of band-pass filter was set at 24 cycles per day; the bandwidth was set at 12cycles per day.

The later is subtracted from the original data and the result indicates an **SES** signal, free from any "zero-level" irregularity. This is presented in the next figure **(10)**.

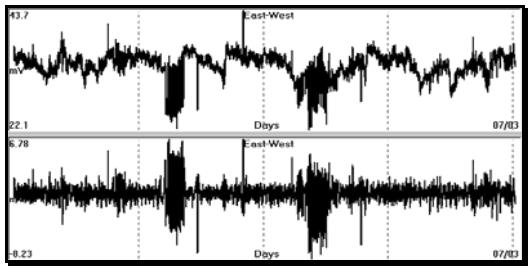

Fig. 10. The upper graph represents the original data, while the lower graph indicates the result of the filtering operation (data recorded by **HIO** monitoring site, (**HIO** 070303 – 070306). The centre of band-pass filter was set at 24 cycles per day and the bandwidth was set at 12cycles per day.

A slightly different approach was followed by Varotsos and Lazaridou (1991), aiming to isolate **SES** signals from a noisy recording. The recordings of two different, in length, "in line" dipoles, were used in a differential mode. The resulted data were drastically "cleaned" from the anthropogenic noise, which was present, in the recording area.

In any case, the tools and the methodologies for any type of data filtering do exist. What must be pointed out is the following: **differential methods act as high-pass filters and therefore, long wavelength components of the processed, electrical signals data, are highly attenuated. <u>This corresponds to a large loss of information useful for the determination of the epicentral area.</u>** The opposite is achieved by integration, which performs as a "low-pass" filter (Thanassoulas, 1991). In this case, **SES** electrical signals, considered, as high frequency components of the registered, electrical field, are totally rejected (considered as noise) from the processed recordings.

**- <u>Magnetotelluric impedance tensor.</u>**

The methodologies which have already been presented are based on the processing of the data itself, with an appropriate filter, after having defined what "**noise**" is and what is "**signal**". In the already presented cases, there is not any reference made, as far as it concerns the origin or the cause of the induced noise. In the specific case, of the noise, induced, through the ionospheric activity, an entirely different methodology was developed: that is the "**Magnetotelluric Impedance Tensor**".

Generally, ionosphere induces currents in the ground. These currents generate potential fields which can be analyzed into its orthogonal components $E_{EW}$ for the **E-W** and $E_{NS}$ for the **N-S** direction. These, ionospherically induced, potentials can be calculated through the use of the **(T)** transfer function, which relates the magnetic fluctuations of the Earth's magnetic field (**Hx, y, z,** components), with the induced on the ground potentials $E_{EW}$ and $E_{NS}$ and the resistivity and tectonic structure of the Earth at the place of the site of investigation.

Chouliaras (1988), presented the details of the methodology and its application on recordings, referred to **VAN** monitoring sites. The procedure, which is followed, is a simple subtraction from the actual recordings which are obtained at the same monitoring site, of the induced "noisy" electric, ionospheric, theoretical components. The resulting output of this procedure is considered, as free, from any ionospheric signal contamination.

## 3. Defining the problem of noise.

The problem of noise contamination of any type of data is generally presented in the following figure **(11)**. The registered data are presented in terms of their frequency spectrum to facilitate the analysis that follows. It is assumed that the signal of interest is present in terms of frequency content in the "**band of interest**", while a wide spectrum of noise spans allover the recorded frequency spectrum. The frequency band which represents the signal of interest is shown in figure **(11)** with right (/), inclined, dashed lines, while the noise spectrum is shown with left (\), inclined, solid lines.

The problem, which is faced, is as follows: how is it possible to extract <u>only the signal of interest</u> from the contaminated, noisy data. The methodologies available, to date are: low-pass filtering, high-pass filtering or their combination in terms of band-pass filtering. The philosophy behind these operations is the same, no matter which methodology is used. A suitable, band-pass filter will be used for the particular case in figure **(11)**.

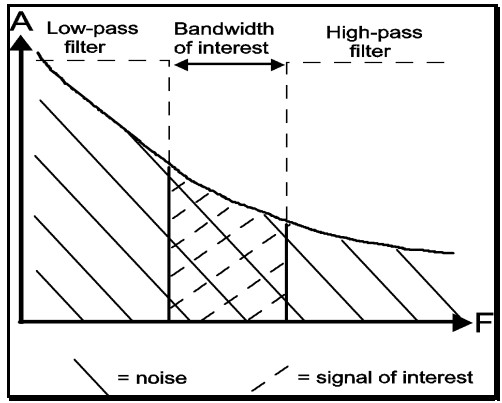

Fig. 11. Schematic presentation of the frequency spectrum of a time series data recording, which has been contaminated by a wide spectrum type of noise.

What is clear, from figure **(11),** is that the application of any band-pass filter upon the data will extract the data of interest, but at the same time it will retain the content part of noise, which coincides with the band of interest. Therefore, the output of any band-pass filter, even of a theoretical one, will retain a certain degree of noise. If the signal to noise ratio (**SNR**) of the original recording is quite large, then there might present no problem at all. In the case when the **SNR** is very small, then the problem becomes serious. The signal of interest could be completely masked by the noise and the entire recording could become, probably, useless.

This problem is usually faced in the recordings of the Earth's electric potential, when very low-level preseismic, electrical signals are to be detected. An example of such a recording is presented in the following figure **(12)**.

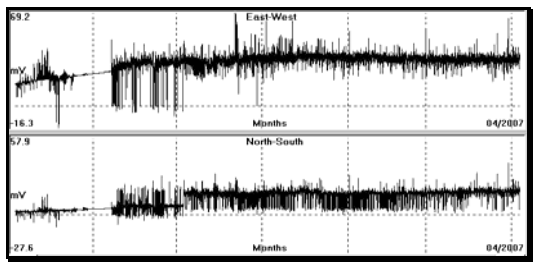

Fig. 12. Presentation of the **EW, NS** components of the Earth's electric potential registered, by **HIO** monitoring site, during the period of 4$^{th}$ October, 2006 to 3$^{rd}$ April, 2007 (**HIO** 061004 – 070403).

A close inspection of the recording indicates that the Earth potential data have been contaminated by noise, which consists mainly of rectangular pulses of very short duration. The application of any of the previously, referred, filters, probably may eliminate the electrical pulses, but at the same time they will affect the data of interest, too. The target of a successful, filtering operation is to eliminate the noise, without affecting the signal of interest, at all. It is evident that a completely different approach must be invented and to be used, as well. A first reaction of the reader could be such as: it is impossible!! Well, is it?

### 4. Theoretical analysis of the "Noise injection filtering methodology".

This methodology, in its principle, although it may seem bizarre, is borrowed by the medical science. Let us assume that a patient is ill, because of an infection by a virus. The pathologist, after having studied the symptoms of the illness, advises the patient to take antibiotics. The level of antibiosis is controlled by the course of the illness. If the patient does not get better, then the level of antibiosis increases, until he becomes totally healthy. The kind of antibiosis depends on the type of virus.

This common, medical procedure will be translated now into the filtering methodology to be applied on a time series data set. The data set corresponds to the patient. The noise contamination corresponds to the illness and the noise type corresponds to the virus. The scientist (physicist, seismologist, geophysicist e.g.) corresponds to the pathologist. What is still missing is the type of antibiosis to be used and its level that will make the data "healthy" again.

The noise, which is observed in figure **(12),** consists of electrical spikes of small duration. Their origin is not important at the present time, but their presence affects badly any preseismic, electrical signal which could be present in this recording. These electrical spikes are considered as **"Dirac delta functions"**, which are rectangular pulses, of an infinitesimal width, and unit area beneath each pulse (Kulhanek, 1976). Each spike can be represented by an infinite number of sinusoidal components of unit amplitude, in phase, at the time of spike occurrence. The later is illustrated in the following figure **(13)**.

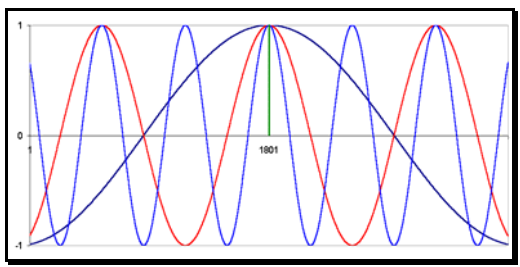

Fig. 13. The "**Dirac delta function**" at time t = 1801 minutes and its first three most significant components.

It is evident, that the frequency components of any electrical spike content in the data set, which fall into the bandwidth of interest, will not be affected at all by the use of any type of filter, which could be used. These spike's frequency components, which are present in the bandwidth of interest, still distort the original signal of interest.

Since the traditional, filtering techniques fail to solve the problem, the medical treatment analogy will be followed (!). What is needed, immediately, is an antibiotic. This is the same, exact "**Dirac delta function**", with opposite sign, which corresponds to a specific, electrical spike. **If a data spike of opposite sign and of the same amplitude is injected in the data series at the same time of the spike occurrence, then the entire frequency noise spectrum, caused by the spike itself only, will be eliminated**. Consequently, only the signals of interest will remain in the band of interest. What is still open in questioning is the amplitude of the injected spike. Since, there is no clue about the noisy spike original amplitude, a range of different amplitudes is used and the resulted, "filtered" data form a "family" of possible, "noise free" registrations.

The entire procedure is demonstrated through the use of a synthetic example. A sine wave was generated, with amplitude of 2 units peak to peak (p-p), which is presented in the following figure **(14)**.

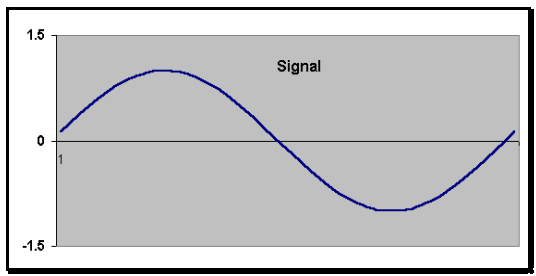

Fig. 14. Schematic presentation of a synthetic sinusoidal signal, with amplitude of 2 units p-p.

The sinusoidal signal of figure **(14)** will be "contaminated" by a noise of about 38 units p-p amplitude. The noise amplitude was generated by a random number generator for the same time span, as for the sinusoidal wave. The form of the calculated noise is shown in the next figure **(15)**.

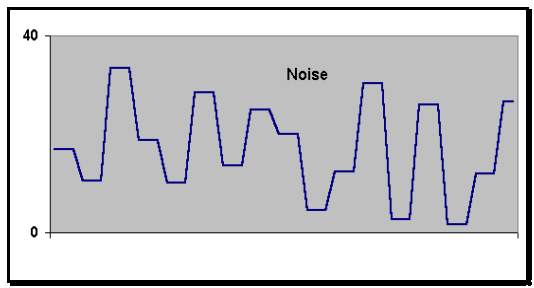

Fig. 15. Presentation of noise, generated for the same time span, as in figure **(14),** by a random procedure.

The **SNR (p-p)** of the sine wave amplitude to the noise amplitude is: **SNR = -26 db.** The sine wave and the noise are combined in one data series, which is presented in the following figure **(16).**

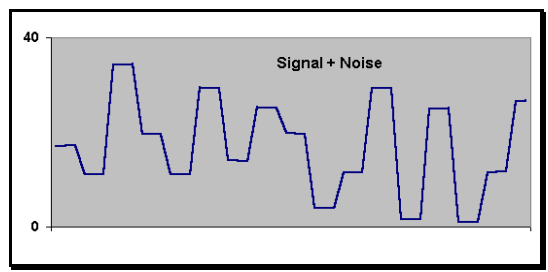

Fig. 16. Synthetic signal generated by a sine wave **(fig. 14)** and random noise **(fig. 15).**

A comparison between the random noise, illustrated in figure **(15)** and the synthetic, noisy data, presented in fig. **(16)** shows, practically, no difference at all. This is caused by the fact that the **SNR** is very small thus, the noise, practically, masks completely any visible sine wave.

Next step towards the recovering of the signal of interest (sine wave) is to apply the methodology, which has already been presented. From the various "**Dirac delta functions**" amplitudes, used, the one with amplitude parameter **(p):**

**p = 1**

generated the best results, shown in the following figure **(17)**

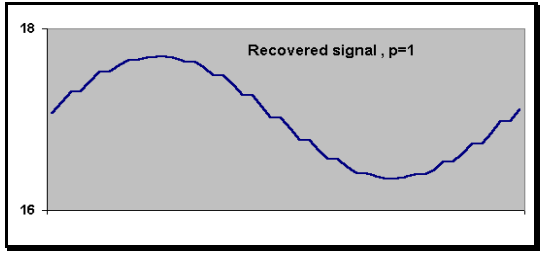

Fig. 17. The recovered signal, from noisy data of figure **(16)** is presented, after the application of the "noise injection methodology".

It is evident that, the recovered signal resembles, pretty well, the original signal, which is used for the purpose of demonstration and refers to the "noise injection methodology".

## 5. Application of the method.

The details of the "noise injection methodology" application are presented as a flow-chart, as follows, in figure (18): It is assumed that a data series (**A**) has been contaminated by a certain amount of spike-like noise. Generally the noise amplitude is not known in advance. Therefore, during the application of this method (**E**) it is required that a "family" (**C**) of noise injection amplitudes is used instead. The later generates a corresponding "family" of filtered data series (**G**). The problem now is the following: which generated filtered data set is the correct one? This is solved in the next step (**F**). At this stage, each filtered data set is compared to a suitable logical criterion (**D**). In case of an agreement between a filtered data set to the used criterion then the specific filtered data set (**B**) is the correct one.

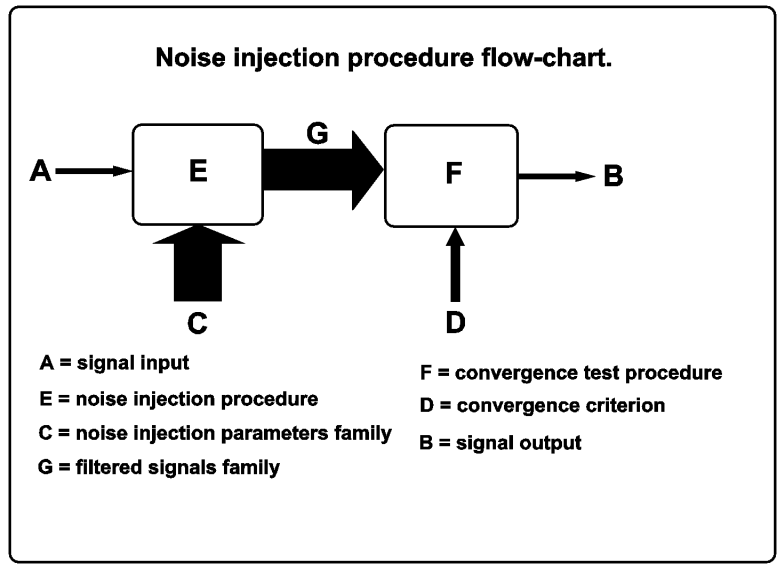

Fig. 18. Flow-chart for the "noise injection" method application. Explanatory details are included in the graph.

Firstly, the procedure (**E**) will be presented as it is applied on a real data set (**A**), registered at **HIO** monitoring site, for a period of six (**6**) months (4$^{th}$ October, 2006 to 3$^{rd}$ April, 2007). The family (**C**) of parameter, called (**p**), used, ranges in the span from **p = .625** to **40**. This parameter controls the amplitude of the injected noise. The filtered examples (family, **G**), presented, refer to both **EW** and **NS** components of the Earth's electric potential. The different (**p**) values, which are used, are also presented in ascending order.

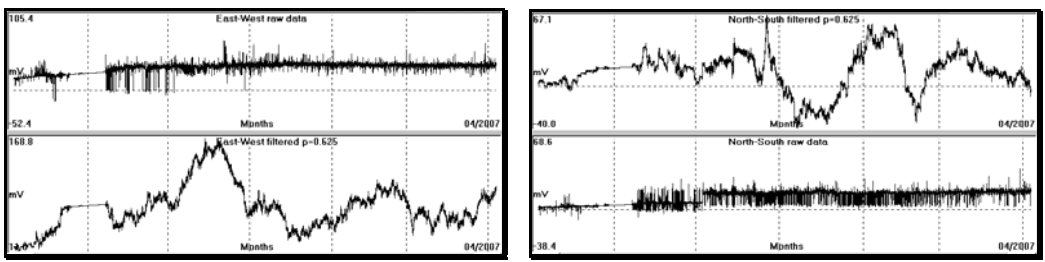

Fig. 19 - 20. **HIO, EW** (left) **and NS** (right) data, processed, by the noise injection methodology, using: **p = 0.625**

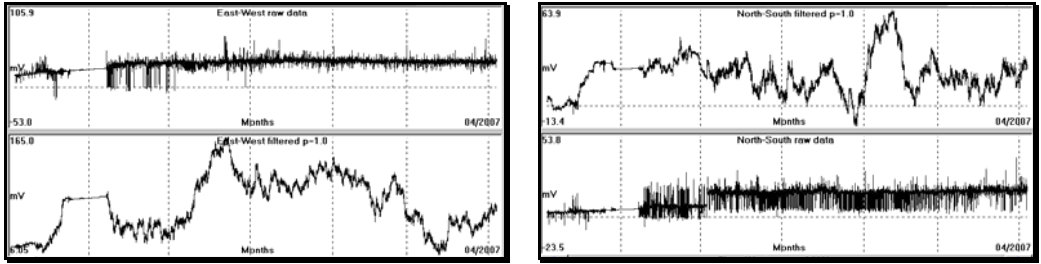

Fig. 21 - 22. **HIO, EW** (left) and **NS** (right) data, processed, by the noise injection methodology, using: **p = 1.0**

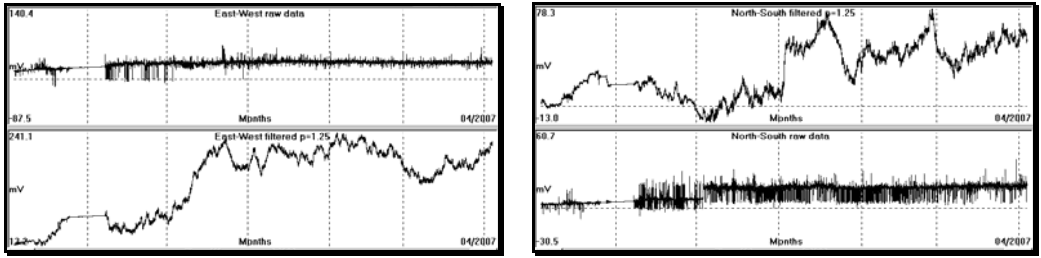

Fig. 23 - 24. **HIO, EW** (left) and **NS** (right) data, processed, by the noise injection methodology, using: **p = 1.25**

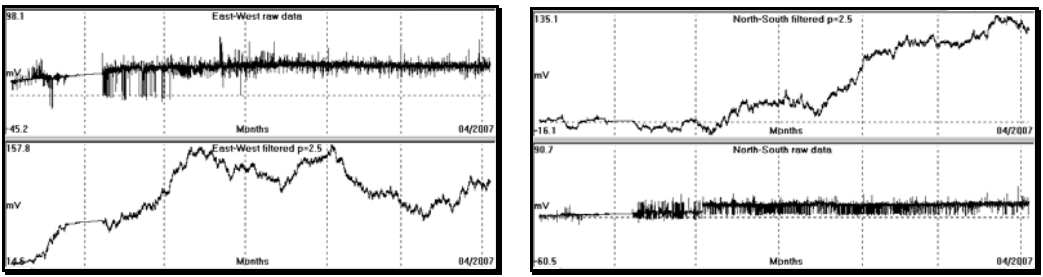

Fig. 25 - 26. **HIO, EW** (left) and **NS** (right) data, processed, by the noise injection methodology, using: **p = 2.5**

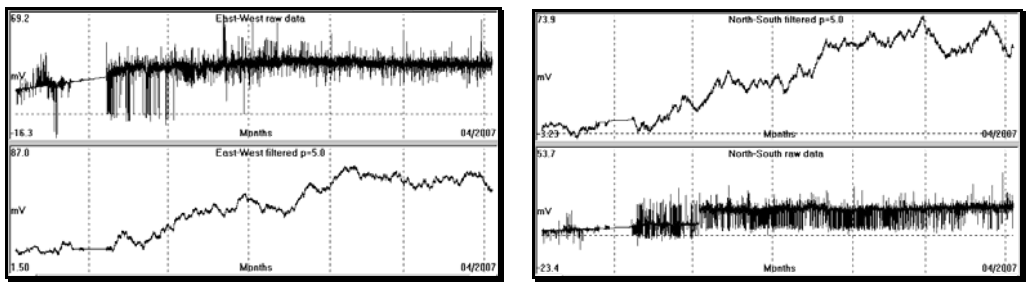

Fig. 27 - 28. **HIO, EW** (left) and **NS** (right) data, processed, by the noise injection methodology, using: **p = 5.0**

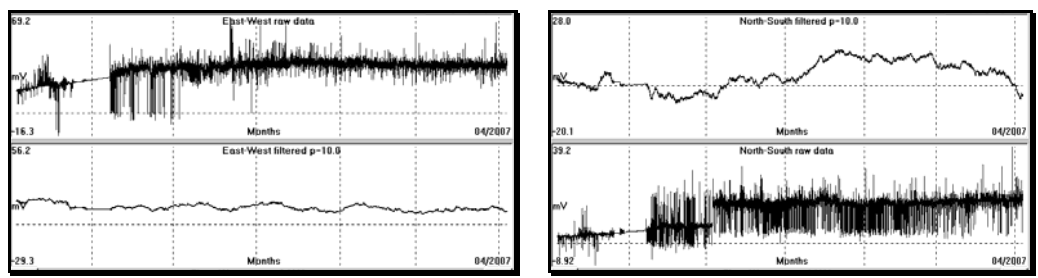

Fig. 29 - 30. **HIO, EW** (left) and **NS** (right) data, processed, by the noise injection methodology, using: **p = 10.0**

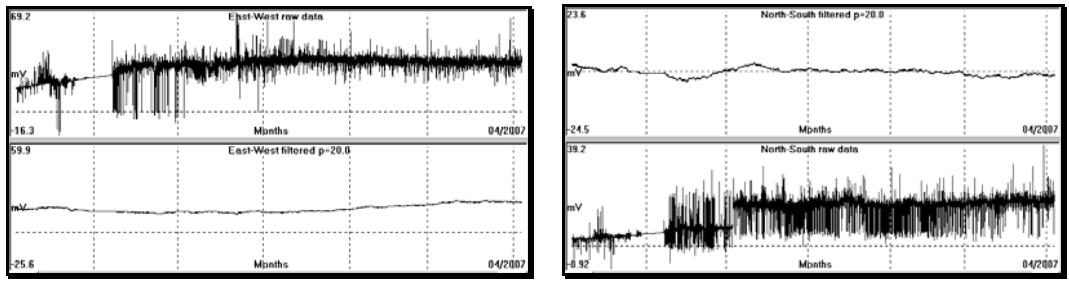

Fig. 31 - 32. **HIO, EW** (left) and **NS** (right) data, processed, by the noise injection methodology, using: **p = 20.0**

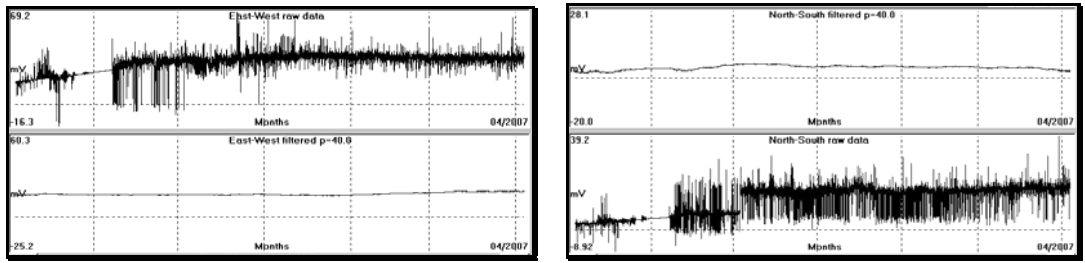

Fig. 33 - 34. **HIO, EW** (left) and **NS** (right) data, processed, by the noise injection methodology, using: **p = 40.0**

A comparison of the previous examples, in relation to the gradual increase of **(p)** parameter value, indicates that for very large **(p)** the noisy data set actually collapses to almost a straight line, in case of absence of any significant, anomalous, electrical signal.

## 6. Application examples on real earthquakes.

Finally, the methodology of "noise injection" was applied on real data, registered, long before strong earthquakes. Three examples will be presented. The first earthquake is the one of **IZMIT**, Turkey (M = 7.5 R, 17$^{th}$ August 1999) and the second one is of **MILOS**, Greece (M = 5.6 R, 21$^{st}$ May 2002). These examples serve the purpose to demonstrate the extraction of the preseismic signal from the noisy data. In these two test examples the convergence criterion (**D**) which was used by the procedure (**F**) was the exact timing (**t**) of the occurred large EQ. A third example (of the Andravida EQ, Ms=7.0R, 8$^{th}$ June, 2008) is used to demonstrate the validity of the obtained preseismic signals. This is achieved by calculating the correct epicentral area by the use of the intersection of the electrical field intensity vectors obtained from two distant monitoring sites (Thanassoulas, 2007).

**Example of application on a real EQ (IZMIT, Turkey EQ, M = 7.5 R, 17$^{th}$ August, 1999).**

The electric potential, generated by **IZMIT** EQ focal area, was recorded (Thanassoulas et al. 2000) by a single dipole (l = 112m, directed almost N-S) at Volos, Greece, at a distance of 650km from the epicentral area. The raw data and the filtered ones are presented in the following figure **(35)**.

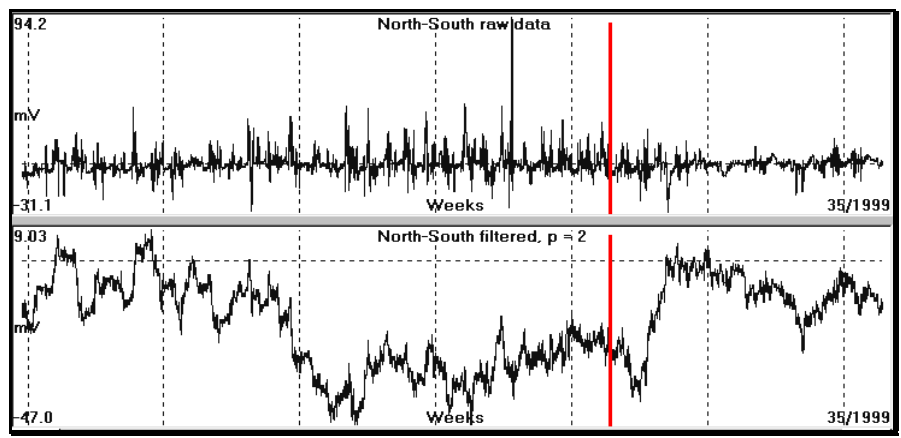

Fig. 35. Raw data (upper graph) and filtered by the "noise injection" methodology (lower graph, **p=2**), which correspond to **IZMIT** earthquake, are, presented. The red bar indicates the time of occurrence of **IZMIT** EQ.

The plateau-type signal (first derivative of the total generated preseismic potential, Thanassoulas, 2007) was recovered pretty well.

**Example of application on a real EQ (MILOS EQ, M = 5.6 R, 21st May 2002).**

The second example, of **MILOS** EQ (M = 5.6 R, 21st May 2002), is presented in the following figure **(36)**. A straight line fits, very well, with the raw data, presented, in the upper graph, while the filtered data (lower graph), is a quite different case.

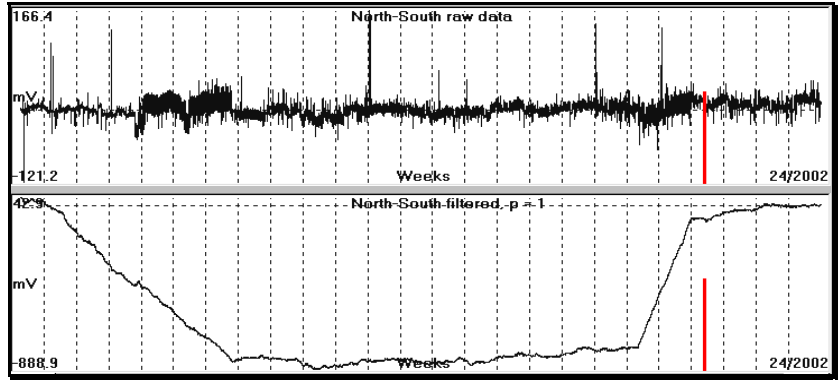

Fig. 36. Raw data (upper graph) and filtered by the noise injection methodology (**p = 1**, lower graph), which correspond to **MILOS** earthquake, are presented. The red bar indicates the time of occurrence of **MILOS** EQ.

In the case of **MILOS** EQ the preseismic electric signal lasted for more than twenty (20) weeks. In this case, too, the preseismic signal has the typical form of the first derivative of the total Earth potential, generated, by the physical mechanism at the focal region. Both, already presented examples, suggest the activation in the focal area of a large-scale piezoelectric mechanism.

**Example of application on Andravida EQ, Greece, (Ms=7.0R, 8th June, 2008).**

In this example two (**2**) data sets are used. The first one was generated by **HIO** monitoring site while the second one was generated by **PYR** monitoring site. In the following figures (**37, 38**) both raw and filtered data along with the best (**p**) used parameter are presented.

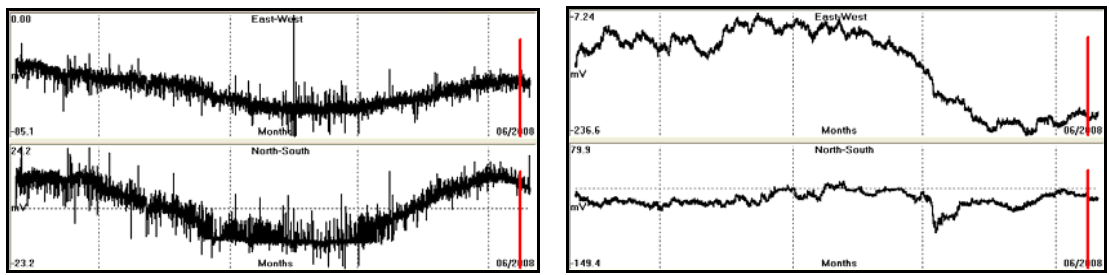

Fig. 37. Left: **HIO** raw data. Right: **HIO** noise injected data, **p = 0.625**

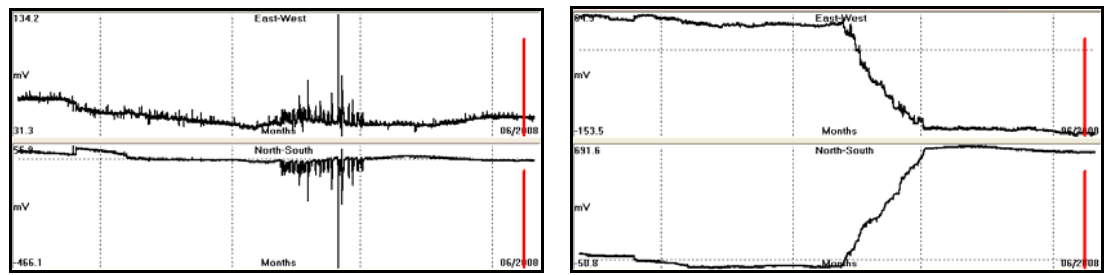

Fig. 38. Left: **PYR** raw data. Right: **PYR** noise injected data, **p = 0.625**

The selection of the (**p**) value as: **p = 0.625** was made by using as convergence criterion (**D**) the true location of the Andavida EQ compared to the numerous intersections which result from the combination of the two different families of azimuthal directions of the earth's electric field intensity vectors. These "family" azimuthal directions were generated from the filtered data "families" from **PYR** and **HIO** monitoring sites (procedure **F**). The details are presented in the subsequent text.

The filtered data were used to construct azimuthal directions (Thanassoulas, 2007) of the preseismic electric signal intensity vector, at each monitoring site. The only one that meets the criterion (**D**) is that with **p = 0.626.** This is depicted in the subsequent figures (**39 - 40**).

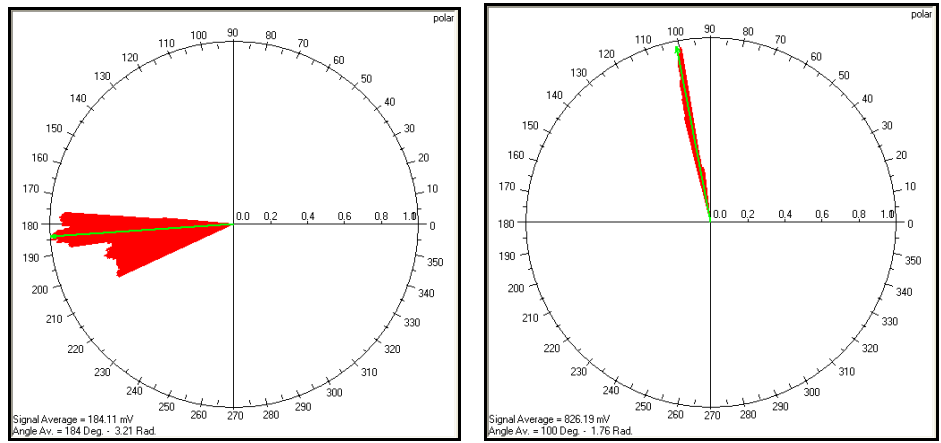

Fig. 39. Azimuthal direction of the preseismic electric field signal intensity vector calculated for **HIO** (left) and **PYR** (right) monitoring sites.

The validation of the selected (**p**) value is made through the calculation of the epicenter area (intersection of **PYR** and **HIO** azimuthal directions of the Earth's electric field intensity vectors) and its comparison to the actual epicenter calculated by ordinary seismological methods. The later is shown in the following figure (**40**).

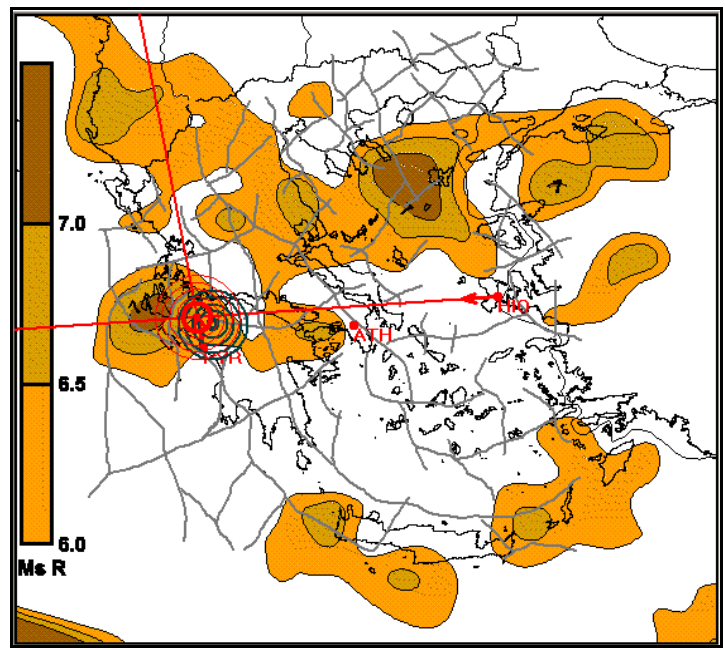

Fig. 40. Actual epicenter of the Andravida EQ (blue circles) calculated by ordinary seismological methods compared to the one (red circle) calculated from the extracted preseismic signal after the application of the "noise injection method" on the original recorded data at **PYR** and **HIO** monitoring sites.

What must be made clear at this point is the fact that the Andravida EQ example is an "a posteriori" one. It was presented just for demonstration purposes of the specific "noise injection" method and by no means is an "earthquake prediction" in terms of its location. The application of this methodology for predictive purposes, in actual conditions, requires three at least, monitoring sites azimuthal directions intersection convergence. This has been analyzed in details by Thanassoulas (2007).

**7. Conclusions.**

The proposed methodology is capable of extracting the preseismic electric signals which are highly contaminated by noise. Since the method is a parametric one using as a basic parameter (**p**) the "noise injection" amplitude, it is necessary to use additional constrains in order to select the correct (**p**) value from the generated "**p**-dependent family" of filtered data series. This is usually achieved by comparing electric field intensity vectors intersections generated from more than three monitoring sites.

In conclusion, methodologies, for the effective rejection of any noise which has contaminated any earthquake precursory, electrical signals, recorded on the ground, at any distance from the epicenter area, do exist. The signal to noise ratio **(SNR),** which is present in these recordings, dictates the most appropriate methodology to be used. When the **SNR** value is quite large, traditional methods of filtering perform equally well. On the contrary, if the **SNR** value is very small then more sophisticated and logical methods (as the presented "noise injection" methodology) must be used in order to extract the desired (if any) preseismic signal.

## 8. References.